\documentclass[aps,twocolumn,prb,showpacs,floatfix,superscriptaddress]{revtex4}
\usepackage{amsfonts}
\usepackage{graphicx}
\usepackage{amsmath}
\usepackage{amssymb}

\begin{document}

\title[Many-body interactions in a quantum wire]{Many-body interactions in a quantum wire in
the integer quantum Hall regime: suppression of exchange-enhanced
$g$ factor}
\author{O. G. Balev}
\email[Electronic address: ]{vbalev@df.ufscar.br}
\affiliation{Departamento de F\'{\i}sica, Universidade Federal de S\~{a}o Carlos,
13565-905, S\~{a}o Carlos, S\~{a}o Paulo, Brazil}
\affiliation{Institute of Semiconductor Physics, NAS of Ukraine, Kiev 03650, Ukraine}
\author{Sanderson Silva}
\affiliation{Departamento de F\'{\i}sica, Universidade Federal de S\~{a}o Carlos,
13565-905, S\~{a}o Carlos, S\~{a}o Paulo, Brazil}
\affiliation{Departamento de F\'{\i}sica, Universidade Federal do Amazonas, Manaus,
Amazonas, Brazil}
\author{Nelson Studart}
\affiliation{Departamento de F\'{\i}sica, Universidade Federal de S\~{a}o Carlos,
13565-905, S\~{a}o Carlos, S\~{a}o Paulo, Brazil}
\keywords{Quantum wires; Quantum Hall liquids, $g$-factor}

\begin{abstract}
The collapse of Hall gaps in the integer quantum Hall liquid in a quantum
wire is investigated. Motivated by recent experiment [Pallecchi \textit{et al}%
. PRB 65, 125303 (2002)] previous approaches are extended to treat
confinement effects and the exchanged enhanced $g$ factor in quantum wires.
Two scenarios for the collapse of the $\nu =1$ state are discussed. In the
first one the $\nu =1$ state becomes unstable at $B_{cr}^{(1)}$, due to the
exchange interaction and correlation effects, coming from the edge-states
screening. In the second scenario, a transition to the $\nu =2$ state occurs
at $B_{cr}^{(2)}$, with a smaller effective channel width, caused by the
redistribution of the charge density. This effect turns the Hartree
interaction essential in calculating the total energy and changes $%
B_{cr}^{(2)}$ drastically. In both scenarios, the exchange enhanced $g$%
-factor is suppressed for magnetic fields lower than $B_{cr}$. Phase
diagrams for the Hall gap collapse are determined. The critical fields,
activation energy, and optical $g$-factor obtained are compared with
experiments. Within the accuracy of the available data, the first scenario
is most probable to be realized.
\end{abstract}

\date{May 20, 2005}
\pacs{73.43.-f, 73.43.Cd, 73.43.Nq, 73.43.Qt}
\maketitle

\section{Introduction}

Effects of electron-electron interaction and lateral confinement in a
two-dimensional electron system (2DES) in the presence of a strong magnetic
field $B$, especially in the integer quantum Hall regime, have attracted
significant attention in recent years. Most of work have been concentrated
on the bulk properties of the quantum Hall liquid. However for wide channels
of the 2DES, the interplay of the edge states and electron-electron
interactions plays dominant role in the understanding of unusual properties
of the Hall liquid. \cite%
{chklovskii92,muller92,dempsey93,gelfand94,balev01,irina02}. Another system
of particular interest, that we are focusing on this paper, is the
quasi-one-dimensional electron system (Q1DES) in narrow channels hereafter
named quantum wires (QW).\cite%
{kinaret90,brey93,suzuki93,wrobel94,balev97,pallecchi02,zhang01,zhang02}.

It is well known that the bulk Land\'{e} $g$-factor that describes the spin
splitting in the presence of the magnetic field is strongly affected by
many-body interactions and the influence should be more drastic as the
dimensionality is reduced. At first glance, the exchange interaction is the
main responsible by the enhancement of the $g$-factor at lower dimensions.
However pronounced effects of quantum confinement and electron correlations
in QW lead to the proposal of different scenarios in order to understand
experimental results.

Kinaret and Lee\cite{kinaret90} found the decreasing of the
exchanged-enhanced spin splitting of a Q1DES in a QW as the width is
reduced. By minimizing the total energy, calculated from the unscreened
exchange interaction for a fixed linear density $n_{L}$, they observed that,
at a certain critical density, the exchange enhancement of the spin
splitting is suppressed. This phase transition occurs because the cost in
the kinetic energy for adding electrons for only one spin-split level
increases up to a critical point in which the population of both spin-split
levels becomes more favorable for the same $n_{L}$.

Only recently the influence of electron correlations coming from the edge
states were taken into account\cite{balev97,zhang01,balev01,zhang02,irina02}%
. It was shown that these effects, associated with the strong screening by
the edge states, are quite relevant both for QWs \cite%
{balev97,zhang01,zhang02} and for wide channels. \cite%
{balev97,balev01,irina02} For the latter system, Balev and Studart\cite%
{balev01} were able to calculate the screened Coulomb potential by
performing an exact infinite summation of a power series in the relevant
parameter $r_{0}=e^{2}/\varepsilon \ell _{0}\hbar \omega _{c}$ which
characterizes the strength of the electron-electron interaction relative to
the cyclotron energy. Here $\ell _{0}=\sqrt{\hbar /m^{\ast }\omega _{c}}$ is
the magnetic length, $\omega _{c}=|e|B/m^{\ast }c$ the cyclotron frequency
and $\varepsilon $ is the background dielectric constant. In the
Balev-Studart (BS) self-consistent nonlocal treatment, the many-particle
energy dispersion relations are obtained from the solution of a \textit{%
single-particle} Schr\"{o}dinger equation determined by an
exchange-correlation potential which is given in terms of the total
single-particle energy and the effective confining potential.\cite{glda} The
latter one is calculated in the self-consistent Hartree approximation (HA)
taking into account screening effects on the external (bare) one-electron
lateral confining potential.\cite{balev01} The BS approach has the merit
that, by considering infinite number of terms in a systematic expansion in
powers of $r_{0}<<1$, its validity is well justified for $r_{0}\lesssim 1$
(in experiments $r_{0}\sim 1$) and for different confining potentials
provided their forms are smooth on the $\ell _{0}$ scale. If we adopt the
parabolic confinement, $V_{y}=m^{\ast }\Omega ^{2}y^{2}/2$, where $\Omega $
is the confinement frequency, that implies $\Omega ^{2}/\omega _{c}^{2}\ll 1$%
. It was shown that edge-state nonlocal correlations change appreciably the
spectrum of the spin-split Landau levels (LLs) leading to a highly
asymmetric Fermi level within the gap between the ($n=0,\sigma =+1)$ and ($%
n=0,\sigma =-1)$ LLs as well the edge group-velocity is drastically
renormalized. As further conclusions, it was shown in BS that the strong
correlation effects induced by the edge states can lead to the collapse of
the fundamental Hall gap, which defines the activation\ $g$-factor. As these
findings are noticeable, we cannot neglect exchange-correlation effects in
calculating many-body properties of the $\nu =1$ quantum Hall liquid in
electron channels.

In this paper, the BS approach is extended to determine the structure of
lowest spin-split LLs ($n=0,$ $\sigma =\pm 1$) in the QW system for the $\nu
=1$ quantum Hall liquid at zero temperature. Strong correlation effects due
to the screening of both left and right edges of the channel are now taken
fully into account. Besides the intrinsic interest on theoretical aspects of
this intensively studied system, our motivation stems mainly from recent
magnetocapacitance experiment in GaAs/AlGaAs QW heterostructures where the
evolution 1D subband filling and spacing was studied as function of
confinement, gate voltage and magnetic field.\cite{pallecchi02} In this
reference, Pallecchi \textit{et al}. compared the experimental results with
the predictions by Kinaret and Lee\cite{kinaret90} and by Balev and
Vasilopoulos\cite{balev97} and concluded that essential improvements of
these models are needed.

We obtain analytically the renormalized, by exchange-correlation effects,
group velocity $v_{g0}^{(1)}$ of the edge states and find that $%
v_{g0}^{(1)}\propto (v_{g0}^{1,H})^{1/2}$, where $v_{g0}^{1,H}$ is the
edge-state group velocity in the HA. We calculate also the enhanced
activation gap $G$, strongly dependent on the exchange-correlation
interaction, for several values of the magnetic field and the Fermi wave
vector $k_{F}$, which defines the width $W$ of the QW because $W$ is
linearly proportional to the number of filled $k$ states for a given band.
Similarly to the wide channel system, we show that the spatial behavior of
the occupied LL in a QW is strongly modified due to electron correlations,
especially near the edges, in comparison with the results in the HA and the
Hartree-Fock approximation (HFA). The position of the Fermi level in the gap
at the centre of the QW is highly asymmetric due to correlation effects
induced by edge states. Though in the HFA, the exchange interaction leads to
the edge-state velocity $v_{g0}^{1,x}$ that diverges logarithmically,
correlation effects restore the smoothness of the single-particle energy as
a function of the oscillator center $y_{0}$, on the $\ell _{0}$ scale.\cite%
{balev97,zhang01,balev01}

In order to understand better the effects of electron-electron interaction
and lateral confinement on the exchange-enhanced spin splitting at the $\nu
=1$ Hall state in QW, we treat two scenarios for the collapse of the
activation gap $G$. In the first one, there is no change in the effective QW
width when the $\nu =1$ state becomes unstable at a critical magnetic field $%
B_{cr}^{(1)}$. As a consequence, no finite redistribution of the electron
charge density occurs at $B_{cr}^{(1)}$, when the Fermi level touches the
bottom of empty LL ($n=0,\sigma =1$). This scenario, proposed in Ref.%
\onlinecite{balev97}, is analyzed here by employing the BS approach to study
the energy spectrum, activation gap and \textquotedblleft
optical\textquotedblright\ $g$-factor of the QW. In the second scenario,
similar to that discussed by Kinaret and Lee,\cite{kinaret90} we include the
very essential Hartree energy, missed in Ref. \onlinecite{kinaret90}. Now
the transition to the $\nu =2$ state happens at a certain $B_{cr}^{(2)}$.
For $B<B_{cr}^{(2)}$, in the $\nu =1$ state, the increase of the kinetic
energy exceeds the energy gain from the exchange energy plus the Hartree
energy, with respect to the $\nu =2$ state, such that the total energy of
the QW in the $\nu =2$ state is lower than in $\nu =1$ state. In this
scenario, the width of the QW becomes two times smaller (if we neglect the
bare $g$-factor $g_{0}$) for $B<B_{cr}^{(2)}$. We make a detailed analysis
of both scenarios and compare their predictions with experimental results%
\cite{wrobel94,pallecchi02} as well as with the predictions of Kinaret and
Lee model.\cite{kinaret90} We show that the first scenario is realized in
experimentally observed collapse of the activation gap of the $\nu =1$
quantum Hall state as in wide channels of Ref. \onlinecite{wrobel94} ($W\sim
3000$ \AA , $\hbar \Omega \sim 0.5$ meV) as for much narrower QWs considered
in recent experiment by Pallecchi \textit{et al}. ($W\sim 500$ \AA , $\hbar
\Omega \sim 5$ meV).\cite{pallechi02}

The outline of our paper is as follows. In Sec. II A, we extend, for a sake
of completeness, on QWs the microscopic formalism of Ref. %
\onlinecite{balev01} for obtaining the screened Coulomb interaction in the
very strong magnetic field limit $r_{0}\ll 1$. In Sec. II B, the BS
approach, for $r_{0}\lesssim 1$, is extended for the quantum Hall liquid in
the QW. We calculate the structure of the LL subband dispersion, the
renormalization, due to exchange and correlations, of the group velocity of
edge states, the activation gap and the optical $g$-factor. The first
scenario of the $\nu =1$ collapse is discussed. In Sec. III A, we revisit
the Kinaret-Lee model and discuss the second scenario for the suppression of
the $\nu =1$ state spin splitting within the HFA,. A detailed comparison of
our results for phase transitions from both scenarios with the experimental
results of Refs. \onlinecite{wrobel94} and \onlinecite{pallecchi02} as well
with those from the model of Ref. \onlinecite{kinaret90} is provided in Sec.
III B. We summarize the key results and present our conclusions in Sec. IV.

\section{Exchange-correlation effects in the quantum wire at $\protect\nu=1$}

We consider the Q1DES in a QW of width $W$ and length $L_{x}=L$ lying in the
$(x,y)$ plane in the presence of a strong magnetic field $B$ pointing up
along the $z$ axis. Choosing the vector potential $\mathbf{A}=-By\widehat{%
\mathbf{x}}$, the single-particle Hamiltonian in the HA is given as $%
\widehat{h}^{0}=[(\widehat{p}_{x}+eBy/c)^{2}+\widehat{p}_{y}^{2}]/2m^{\ast
}+V_{y}+g_{0}\mu _{B}\widehat{\sigma }_{z}B/2$, where the confining
potential $V_{y}=m^{\ast }\Omega ^{2}y^{2}/2$, $g_{0}$ is the bare Land\'{e}
g-factor, $\mu _{B}$ the Bohr magneton, and $\hat{\sigma}_{z}$ the $z$%
-component Pauli matrix. The eigenvalues and eigenfunctions are given by $%
\epsilon _{n,k_{x},\sigma }=(n+1/2)\hbar \tilde{\omega}+\hbar ^{2}k_{x}^{2}/2%
\tilde{m}+\sigma g_{0}\mu _{B}B/2$ and $\psi _{nk_{x}\sigma }(\mathbf{r}%
,\sigma _{1})=\left\langle \mathbf{r}|nk_{x}\rangle |\sigma \right\rangle $,
with $\left\langle \mathbf{r}|nk_{x}\right\rangle =\exp (ik_{x}x)\Psi
_{n}(y-y_{0}(k_{x}))/\sqrt{L}$ and spin function $|\sigma \rangle =\psi
_{\sigma }(\sigma _{1})=\delta _{\sigma \sigma _{1}}$, $\sigma _{1}=\pm 1$.
Here $\tilde{\omega}=(\omega _{c}^{2}+\Omega ^{2})^{1/2}$, $\tilde{m}%
=m^{\ast }\tilde{\omega}^{2}/\Omega ^{2}$, $y_{0}(k_{x})=\hbar \omega
_{c}k_{x}/m^{\ast }\tilde{\omega}^{2}$, $\Psi _{n}(y)$ is a harmonic
oscillator function.

For the $\nu =1$ quantum Hall liquid in the HA the right (left) edge of the
occupied $(n=0,\sigma =1)$ LL is denoted by $y_{r0}^{(1)}=\hbar \omega
_{c}k_{F}/m^{\ast }\tilde{\omega}^{2}$ ($-y_{r0}^{(1)}$), where $k_{F}=(%
\tilde{\omega}/\hbar \Omega )\sqrt{2m^{\ast }\Delta _{F0}^{(1)}}$ is the
Fermi vector; this level is occupied only for $|k_{x}|\leq k_{F}$, $\Delta
_{F0}^{(1)}=E_{F}^{H}-\hbar \tilde{\omega}/2-g_{0}\mu _{B}B/2$, and $%
E_{F}^{H}$ is the Fermi energy in the HA. The QW width is $W=2y_{r0}^{(1)}$.
The group velocity of the edge states in the HA, at the right (left) edge of
the QW is given by $v_{g0}^{1,H}=\partial \epsilon _{0,k_{F},1}/\hbar
\partial k_{x}=\hbar k_{F}/\tilde{m}$ $(-\hbar k_{F}/\tilde{m}).$

For the parabolic confinement, the essential matrix elements were evaluated
in Ref.\onlinecite{balev97}. The result is

\begin{align}
<n^{^{\prime}}k_{x}^{^{\prime}}|e^{i\mathbf{q}\cdot\mathbf{r}}|nk_{x}> &
=(n^{^{\prime}}k_{x}^{^{\prime}}|e^{iq_{y}y}|nk_{x})\delta_{q_{x},-k_{-}}
\notag \\
& =(\frac{n^{\prime}!}{n!})^{1/2}(\frac{aq_{x}+iq_{y}}{\sqrt{2}/\tilde{\ell}}%
)^{m}e^{-u/2}  \notag \\
& \times L_{n^{^{\prime}}}^{m}(u)e^{iaq_{y}k_{+}\tilde{\ell}%
^{2}/2}\delta_{q_{x},-k_{-}},  \label{1}
\end{align}
where $k_{\pm}=k_{x}\pm k_{x}^{^{\prime}}$, $m=n-n^{^{\prime}}$,$\ a=\omega
_{c}/\tilde{\omega}$,$\ u=[a^{2}q_{x}^{2}+q_{y}^{2}]\tilde{\ell}^{2}/2$,$\
\tilde{\ell}=(\hbar/m^{\ast}\tilde{\omega})^{1/2}$ is the renormalized
magnetic length, and $L_{n^{^{\prime}}}^{m}(u)$ the Laguerre polynomial.
Observe that the Eq. (7) of Ref.\onlinecite{kinaret90} differs from Eq. (\ref%
{1}) and the $\mathbf{q}$-anisotropy of $<n^{^{\prime}}k_{x}^{^{\prime}}|e^{i%
\mathbf{q}\cdot\mathbf{r}}|nk_{x}>$ should be pointed out especially for $%
\omega_{c}/\Omega\alt1$.

\subsection{Many-body interactions for $r_{0} \ll 1$}

We will now consider exchange-correlation effects in the QW for the strong
magnetic field limit, $r_{0}\ll1$, when only the lowest spin-up ($\sigma=+1$%
) LL is occupied. The exchange and correlation contributions to the
single-particle energy of this LL $E_{0,k_{x},1}=\epsilon_{0,k_{x},1}+%
\epsilon_{0,k_{x},1}^{xc}$ in the screened Hartree-Fock approximation (SHFA)
is given as\cite{balev97}

\begin{align}
\epsilon_{0,k_{x},1}^{xc} & =-\frac{1}{8\pi^{3}}%
\int_{-k_{F}}^{k_{F}}dk_{x}^{^{\prime}}\int_{-\infty}^{\infty}dq_{y}\int_{-%
\infty}^{\infty}dq_{y}^{^{\prime}}V^{s}(k_{-},q_{y};q_{y}^{^{\prime}})
\notag \\
&
\times(0k_{x}|e^{iq_{y}y}|0k_{x}^{^{\prime}})(0k_{x}^{^{%
\prime}}|e^{iq_{y}^{^{\prime}}y}|0k_{x}),  \label{2}
\end{align}
where $V^{s}(q_{x},q_{y};q_{y}^{^{\prime}})$ is the Fourier transform of the
screened Coulomb interaction which can be evaluated within the random phase
approximation (RPA). For comparison with experiments,\cite%
{wrobel94,pallecchi02} we will assume further in Sec. II that $%
\Omega^{2}/\omega_{c}^{2}\ll1$, then $\tilde{\ell}\approx\ell_{0}=(\hbar/m^{%
\ast}\omega_{c})^{1/2}$.

In order to calculate the screened Coulomb interaction, we follow closely
Ref. \onlinecite{balev01}. The integral equation for the Fourier components
of the induced charge density, by the test electron charge located at ($%
x_{0},y_{0}$), is given as%
\begin{align}
\rho (q_{x},y;y_{0})& =-r_{1}^{H}\;\sum_{k=0}^{1}\Pi
_{00}(y,(-1)^{k}k_{F},(-1)^{k}k_{F}-q_{x})  \notag \\
& \times \int_{-\infty }^{\infty }d\tilde{y}\int_{-\infty }^{\infty
}dy^{\prime }\ K_{0}(|q_{x}||\tilde{y}-y^{\prime }|)  \notag \\
& \times \Pi _{00}(\tilde{y},(-1)^{k}k_{F},(-1)^{k}k_{F}-q_{x})  \notag \\
& \times \lbrack \rho (q_{x},y^{\prime };y_{0})+e\delta (y^{\prime }-y_{0})],
\label{3}
\end{align}%
where $k=0$ $(1)$ term is related to the contribution coming from the right
(left) edge states, $r_{1}^{H}=e^{2}/(\pi \hbar \varepsilon v_{g0}^{1,H})$
is a characteristic dimensionless parameter, $\Pi _{n_{\alpha }n_{\beta
}}(y,k_{x\alpha },k_{x\beta })=\Psi _{n_{\alpha }}[y-y_{0}(k_{x\alpha
})]\Psi _{n_{\beta }}[y-y_{0}(k_{x\beta })]$, and $K_{0}(x)$ is the modified
Bessel function. Equation (\ref{3}) is similar to Eq. (4) of Ref. \onlinecite%
{balev01}. However, notice the essential point that two terms are now
present in Eq. (\ref{3}) because one cannot neglect contributions from both
left and right edges at any point of the QW.

We look for a solution of Eq. (\ref{3}) in the form%
\begin{equation}
\rho (q_{x},y;y_{0})=\sum_{k=0}^{1}\rho ^{(k)}\;\Pi _{00}[\tilde{y}%
,(-1)^{k}k_{F},(-1)^{k}k_{F}-q_{x}],  \label{4}
\end{equation}%
where $\rho ^{(k)}\equiv \rho ^{(k)}(q_{x},y_{0})$. Substituting the Eq. (%
\ref{4}) into Eq. (\ref{3}), we obtain the system of two linear
inhomogeneous equations with respect to $\rho ^{(k)}(q_{x},y_{0})$, for $%
k=0,1$. Calculating these functions and then taking the Fourier transform $%
\rho (q_{x},q_{y};q_{y}^{\prime })$ of $\rho (q_{x},y;y_{0})$ it follows
from $V^{s}(q_{x},q_{y};q_{y}^{\prime })=(2\pi e/\varepsilon \sqrt{%
q_{x}^{2}+q_{y}^{2}})[2\pi e\delta (q_{y}+q_{y}^{\prime })+\rho
(q_{x},q_{y};q_{y}^{\prime })]$ that%
\begin{align}
V^{s}(q_{x},q_{y};q_{y}^{\prime })& =\frac{4\pi ^{2}e^{2}}{\varepsilon \sqrt{%
q_{x}^{2}+q_{y}^{2}}}{\LARGE \{}\delta (q_{y}+q_{y}^{\prime })  \notag \\
& -\frac{r_{1}^{H}\exp [iq_{x}(q_{y}+q_{y}^{\prime })\ell _{0}^{2}/2]}{%
\Delta ^{H}(q_{x})\sqrt{q_{x}^{2}+(q_{y}^{\prime })^{2}}}  \notag \\
& \times e^{-[2q_{x}^{2}+q_{y}^{2}+(q_{y}^{\prime })^{2}]\ell
_{0}^{2}/4}[(1+r_{1}^{H}M(0,q_{x}))  \notag \\
& \times \cos (k_{F}(q_{y}+q_{y}^{\prime })\ell
_{0}^{2})-r_{1}^{H}M(2k_{F},q_{x})  \notag \\
& \times \cos (k_{F}(q_{y}-q_{y}^{\prime })\ell _{0}^{2})]{\LARGE \}},
\label{5}
\end{align}%
where%
\begin{equation}
M(k_{x},q_{x})=e^{-q_{x}^{2}\ell _{0}^{2}/2}\int_{0}^{\infty }dq_{y}\frac{%
e^{-q_{y}^{2}\ell _{0}^{2}/2}}{\sqrt{q_{x}^{2}+q_{y}^{2}}}\cos
[q_{y}k_{x}\ell _{0}^{2}],  \label{6}
\end{equation}%
and%
\begin{equation}
\Delta
^{H}(q_{x})=[1+r_{1}^{H}M(0,q_{x})]^{2}-[r_{1}^{H}M(2k_{F},q_{x})]^{2}.
\label{7}
\end{equation}%
From Eq. (\ref{6}) it follows that $M(0,q_{x})=2^{-1}\exp (-q_{x}^{2}\ell
_{0}^{2}/4)K_{0}(q_{x}^{2}\ell _{0}^{2}/4)$ and for $2k_{F}\ell _{0}\gg 1$, $%
M(2k_{F},q_{x})\approx K_{0}(2k_{F}q_{x}\ell _{0}^{2})$. The first term in
the curly brackets of Eq. (\ref{5}) is the bare Coulomb interaction which
leads to the exchange contribution\cite{kinaret90,balev97,dempsey93}, with
neglected small corrections of the order of $\Omega ^{2}/\omega _{c}^{2}\ll
1 $.

Substituting Eq. (\ref{5}) into Eq. (\ref{2}), we obtain the single-particle
exchange-correlation energy as

\begin{align}
\epsilon_{0,k_{x},1}^{xc} & =-\frac{e^{2}}{\pi\varepsilon}%
\int_{-k_{F}}^{k_{F}}\frac{dk_{x}^{\prime}}{\Delta^{H}(k_{x}-k_{x}^{\prime})}%
{\LARGE \{}M(0,k_{x}-k_{x}^{\prime})  \notag \\
&
\times\Delta^{H}(k_{x}-k_{x}^{%
\prime})-r_{1}^{H}[(1+r_{1}^{H}M(0,k_{x}-k_{x}^{\prime}))  \notag \\
& \times\sum_{k=0}^{1}M^{2}(k_{x}+(-1)^{k}k_{F},k_{x}-k_{x}^{\prime })
\notag \\
& -2r_{1}^{H}M(2k_{F},k_{x}-k_{x}^{\prime})M(k_{x}-k_{F},k_{x}-k_{x}^{\prime
})  \notag \\
& \times M(k_{x}+k_{F},k_{x}-k_{x}^{\prime})]{\LARGE \}}.  \label{8}
\end{align}
The first term in the curly brackets of Eq. (\ref{8}) gives the exchange
energy. Remaining terms are the important electron-correlation contributions
to the energy coming from the edge-states screening of both left and right
edges of the QW.

\subsection{Structure of Landau level subbands for $r_{0} \alt 1$}

In order to make comparisons between theoretical predictions and the results
of actual experiments, it is necessary to go beyond the strong magnetic
field limit ($r_{0}\ll 1)$, considered in Sec II A, to reach the regime
achieved experimentally ($r_{0}\sim 1).$ We begin by defining a new
characteristic dimensionless parameter $r_{1}=e^{2}/(\pi \hbar \varepsilon
v_{g0}^{(1)})$ instead of $r_{1}^{H}$ and assuming that the approximation is
still valid for $r_{0}\lesssim 1.$ Then the total single-particle energy of
the ($n=0,\sigma =1$) LL $E_{0,k_{x},1}=\epsilon _{0,k_{x},1}+\epsilon
_{0,k_{x},1}^{xc}$, where $\epsilon _{0,k_{x},1}^{xc}$ is given by Eq. (\ref%
{8}), is given by%
\begin{align}
E_{0,k_{x},1}& =\frac{\hbar \omega _{c}}{2}-\frac{|g_{0}|\mu _{B}B}{2}+\frac{%
m^{\ast }\Omega ^{2}\ell _{0}^{4}}{2}k_{x}^{2}-\frac{e^{2}}{\pi \varepsilon }%
\   \notag \\
& \times \int_{-k_{F}-k_{x}}^{k_{F}-k_{x}}\frac{dx}{\Delta (x_{\delta })}%
\{M(0,x_{\delta })\Delta (x_{\delta })\   \notag \\
& -r_{1}[(1+r_{1}M(0,x_{\delta }))(M^{2}(k_{x}-k_{F},x_{\delta })\   \notag
\\
& +M^{2}(k_{x}+k_{F},x_{\delta }))-2r_{1}M(2k_{F},x_{\delta })\   \notag \\
& \times M(k_{x}-k_{F},x_{\delta })M(k_{x}+k_{F},x_{\delta })]\},  \label{9}
\end{align}%
where $x_{\delta }=\sqrt{x^{2}+\delta ^{2}/\ell _{0}^{2}}$, $\Delta (x)$ it
follows from $\Delta ^{H}(x)$ after changing $r_{1}^{H}$ by $r_{1}\equiv
r_{1}(v_{g0}^{(1)})$. The renormalized group velocity of the edge states is
defined from Eq. (\ref{9}) as $v_{g0}^{(1)}=(\partial E_{0,k_{x},1}/\hbar
\partial k_{x})_{k_{x}=k_{F}}$. This is the condition of self-consistency in
the BS approach for the QW. Renormalized by exchange-correlation effects, $%
v_{g0}^{(1)}$ is given by a positive solution of the cubic equation%
\begin{align}
\tilde{v}_{g}^{3}& +(M(0,\delta /\ell _{0})-\tilde{v}_{g}^{H})\tilde{v}%
_{g}^{2}-2\tilde{v}_{g}^{H}M(0,\delta /\ell _{0})\tilde{v}_{g}  \notag \\
& -\tilde{v}_{g}^{H}[M^{2}(0,\delta /\ell _{0})-M^{2}(2k_{F},\delta /\ell
_{0})]=0,  \label{10}
\end{align}%
where $\tilde{v}_{g}=1/r_{1}$, $\tilde{v}_{g}^{H}=1/r_{1}^{H}$. This
equation was calculated by using $[\partial M(k_{x}-k_{F},x_{\delta
})/\partial k_{x}]_{k_{x}=k_{F}}=0$ and for assumed restriction $2k_{F}\ell
_{0}\gg 1$. A small parameter $\delta \ll 1$ was introduced in order to
avoid the weak logarithmic divergence for $x\rightarrow 0$. Here $M(0,\delta
/\ell _{0})\approx \lbrack \ln (2\sqrt{2}/\delta )-\gamma /2]$ and $%
M(2k_{F},\delta /\ell _{0})\approx K_{0}(2k_{F}\ell _{0}\delta )\approx
\lbrack \ln (1/k_{F}\ell _{0}\delta )-\gamma ]$, where $\gamma $ is the
Euler constant and $2k_{F}\ell _{0}\delta \ll 1$. It is worth to point out
that by formally discarding the terms containing $M^{2}(2k_{F},\delta /\ell
_{0})$ of Eq. (\ref{10}), which corresponds to neglect the correlations due
to left edge-states of the QW, the Eq. (14) of Ref. \onlinecite{balev01} is
obtained. From Eq. (\ref{10}) for the condition $\tilde{v}_{g}^{H}\ll
\lbrack \ln (8k_{F}^{2}\ell _{0}^{2})+\gamma ]$, which is well satisfied for
the assumptions made, we find that only one root%
\begin{align}
v_{g0}^{(1)}& =\sqrt{\frac{e^{2}}{\pi \hbar \varepsilon }v_{g0}^{1,H}}%
\{[M^{2}(0,\delta /\ell _{0})-M^{2}(2k_{F},\delta /\ell _{0})]  \notag \\
& \;\;\;\;\;\;\;\;\;\;\;\;\;\;\;\;\;\;\;\;\;\times M^{-1}(0,\delta /\ell
_{0})\}^{1/2}+v_{g0}^{1,H}\   \notag \\
& \approx \sqrt{\frac{e^{2}}{\pi \hbar \varepsilon }v_{g0}^{1,H}}[\ln
(8k_{F}^{2}\ell _{0}^{2})+\gamma ]^{1/2}+v_{g0}^{1,H},  \label{11}
\end{align}%
satisfies the physical requirement of $v_{g0}^{(1)}\geq 0$, i.e., the
occupied LL is below $E_{F}$\ for $k_{x}$\ within the interval $%
(-k_{F},k_{F})$. From Eq. (\ref{11}) it follows that $%
v_{g0}^{(1)}/v_{g0}^{1,H}\approx \lbrack r_{1}^{H}(\ln (8k_{F}^{2}\ell
_{0}^{2})+\gamma )]^{1/2}\gg 1$ and we finally obtain $v_{g0}^{(1)}\propto
\sqrt{v_{g0}^{1,H}}$. Note that the approximate expression in Eq. (\ref{11}%
), is independent of the small parameter $\delta $, contrary to the result
obtained for wide channels.\cite{balev01} \ Our Eq. (\ref{11}) is
essentially different from Eq. (21) of Ref. \onlinecite{zhang01} that gives $%
v_{g0}^{(1)}\approx v_{g0}^{1,H}$, for $r_{0}\leq 1$. If we apply Eq. (\ref%
{11}) to the actual parameters of samples 1 and 2 of Ref. %
\onlinecite{wrobel94} it follows that $v_{g0}^{(1)}/v_{g0}^{1,H}\approx 10.4$
($9.1$) and $19.6$ ($17.0$) for $\delta \rightarrow 0$ ($\delta =10^{-3}$),
respectively. In Ref. \onlinecite{balev97} it was found $%
v_{g0}^{(1)}/v_{g0}^{1,H}\approx 5$ and $10$ for samples 1 and 2, at $\delta
=10^{-3}$. On the other hand, these ratios were calculated numerically by a
weighted iterative method in Ref. \onlinecite{zhang01} and the values
obtained are $v_{g0}^{(1)}/v_{g0}^{1,H}\approx 6.9$ and $11$ which are close
to our results and in contrast with the analytical result $%
v_{g0}^{(1)}/v_{g0}^{1,H}\approx 1.0$, for $r_{0}\leq 1$, given in Refs. %
\onlinecite{zhang01} and \onlinecite{zhang02}. The last line in Eq. (\ref{11}%
) can be rewritten as $v_{g0}^{(1)}/v_{g0}^{1,H}=\sqrt{r_{0}/\pi }(\omega
_{c}/\Omega )\{[\ln (8k_{F}^{2}\ell _{0}^{2})+\gamma ]/k_{F}\ell
_{0}\}^{1/2}+1$, for a parabolic $V_{y}$. However, Eqs. (\ref{10}) and (\ref%
{11}) are valid for any confining potential $V_{y}$ that satisfies the
condition of smoothness on $\ell _{0}$ scale. For instance we can assume
here large variation of $v_{g0}^{1,H}$ for a fixed $W=2k_{F}\ell _{0}^{2}$.
In particular, $v_{g0}^{1,H}$ can approach zero due to the flattening effect%
\cite{chklovskii92}, while $W/\ell _{0}=2k_{F}\ell _{0}\gg 1$ is kept
constant.

The activation gap, defined by the energy difference between the bottom of ($%
n=0,\sigma =-1$) LL and the Fermi level, is given by $%
G(v_{g0}^{1,H})=E_{0,0,-1}-E_{0,k_{F},1}=\epsilon _{0,0,-1}-E_{0,k_{F},1}$,
where the Fermi level, $E_{F}=E_{0,k_{F},1}$, follows from Eqs. (\ref{9})-(%
\ref{11}). The result is%
\begin{align}
G& =|g_{0}|\mu _{B}B-\frac{m^{\ast }\omega _{c}^{2}}{2\Omega ^{2}}%
(v_{g0}^{1,H})^{2}+\frac{e^{2}}{\pi \varepsilon \ell _{0}}%
\int_{0}^{2k_{F}\ell _{0}}du  \notag \\
& \times {\LARGE \{}M(0,u_{\delta }/\ell _{0})[1+R_{1}M(0,u_{\delta }/\ell
_{0})]-R_{1}  \notag \\
& \times M^{2}(2k_{F},u_{\delta }/\ell _{0}){\LARGE \}}{\LARGE \{}%
[1+R_{1}M(0,u_{\delta }/\ell _{0})]^{2}  \notag \\
& -R_{1}^{2}M^{2}(2k_{F},u_{\delta }/\ell _{0}){\LARGE \}}^{-1},  \label{12}
\end{align}%
where $R_{1}\equiv R_{1}(v_{g0}^{1,H})$ is the function obtained from $%
r_{1}(v_{g0}^{(1)})$, after using the solution $%
v_{g0}^{(1)}=v_{g0}^{(1)}(v_{g0}^{1,H})$ of Eq. (\ref{10}) and $u_{\delta }=%
\sqrt{u^{2}+\delta ^{2}}$. Notice that due to the terms depending on $%
M(2k_{F},\delta /\ell _{0})$, the Eq. (\ref{12}) is essentially different
from Eq. (16) of Ref. \onlinecite{balev01}. We emphasize that the edge-state
correlation effects constrain the Fermi level of the interacting system at
the centre of the QW to be much more closer to the bottom of the empty ($%
n=0,\sigma =-1$) LL than to the bottom of the occupied ($n=0,\sigma =1$) LL.
Then we must say that $G$ is the actual activation gap of the QW.
\begin{figure}[tbp]
\includegraphics*[width=1.0\linewidth]{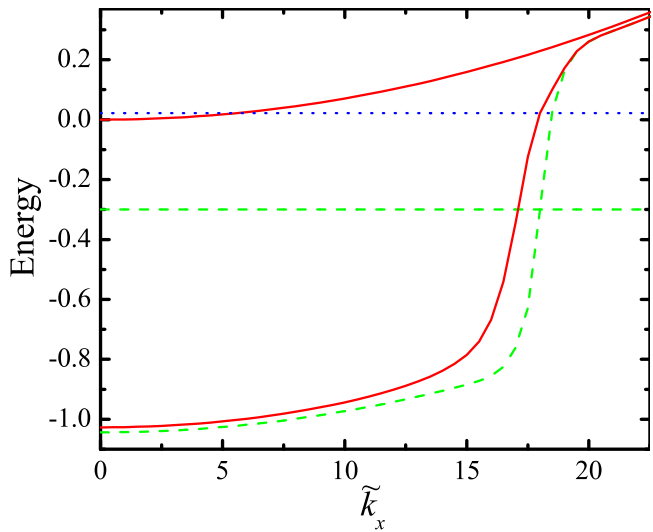}
\caption{(Color on line) Energy dispersion curves for the lowest levels of
the quantum wire (units of $\hbar \protect\omega _{c}$) as a function of $%
\tilde{k}_{x}=k_{x}\ell _{0}$. The bottom (top) red solid line represents $%
E_{0,k_{x},1}$ ($E_{0,k_{x},-1}$) and the blue dot horizontal line
gives the exact position of $E_{F}$, when exchange-correlation
effects are taken into account in the BS approach. The green
dashed curve shows $E_{0,k_{x},1}$ obtained within the HFA. The
horizontal green dashed line indicates the
position of the Fermi level $E_{F}$ within the HFA. The used parameters are $%
B=10$ T, $\hbar \Omega =0.65$ meV, and $k_{F}\ell _{0}=18.0$ ($W\approx 0.29$
$\protect\mu $m), which correspond to parameters of sample 1 in Ref.
\onlinecite{wrobel94}. Here $r_{0}\approx 0.82$, $\protect\omega _{c}/\Omega
\approx 26.6$, $\protect\delta =10^{-3}$, and $v_{g}/v_{g}^{H}\approx 8.5$.
As for these conditions the activation gap is negative ($G_{a}\approx -2.9$%
), there is no any stable $\protect\nu =1$ QHE state, in agreement
with experiment\protect\cite{wrobel94} and, respectively, here the
blue dot line actually indicates the quasi-Fermi level.}
\label{Fig1}
\end{figure}

In order to assess the effect of many-body interactions on $G$, we define a
dimensionless activation gap as $G_{a}(v_{g0}^{1,H})=G/(|g_{0}|\mu _{B}B/2)$.%
\cite{balev01} In the absence of many-body interactions, i.e., in HA $%
G_{a\max }\equiv G_{a\max }^{H}=1$. Then the activation gap is enhanced when
$G_{a}>1$. Indeed $G_{a}$ can be understood as the activation $g$-factor of
the QW given in units of the bare $g$-factor $g_{0}$. We see that a critical
magnetic field $B_{cr}^{(1)}$ is achieved when $G_{a}=0.$ For $%
B>B_{cr}^{(1)} $ we have $G_{a}>0$ and the $\nu =1$ state is
thermodynamically stable. Otherwise for $B<B_{cr}^{(1)}$, $G_{a}<0$ and the $%
\nu =1$ state is unstable. Furthermore, Pallecchi \textit{et al.}\cite%
{pallecchi02} were able to obtain some average value of the effective,
spatially inhomogeneous, \textquotedblleft optical\textquotedblright\ $g$%
-factor $g_{op}^{\ast }=(E_{0,k_{x},-1}-E_{0,k_{x},1})/\mu _{B}B$. Notice,
in agreement with experiment, that $g_{op}^{\ast }(k_{x})$ can be large, due
to the exchange enhancement, even when $G_{a}$ goes to $0$.
\begin{figure}[tbp]
\includegraphics*[width=1.0\linewidth]{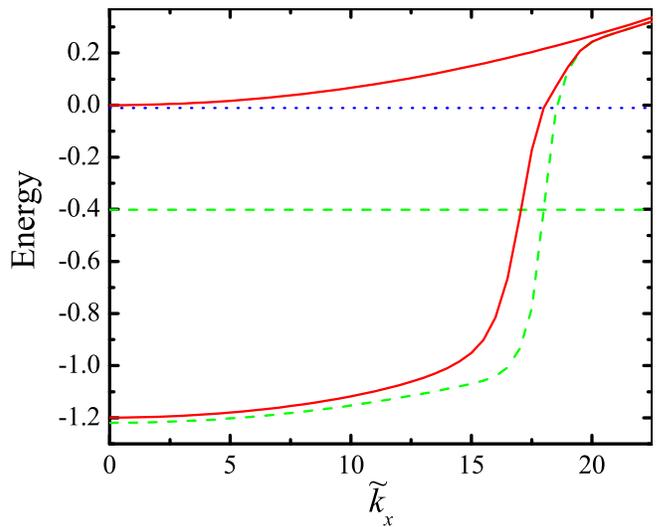}
\caption{(Color on line) Same as in Fig. 1 for the parameters pertinent to
sample 2 of Ref. \onlinecite{wrobel94}: $B=7.3$ T, $\hbar \Omega =0.46$ meV,
and $k_{F}\ell _{0}=18.0$ ($W\approx 0.34$ $\protect\mu $m). Now $%
r_{0}\approx 0.96$, $\protect\omega _{c}/\Omega \approx 27.4,$ $%
v_{g}/v_{g}^{H}\approx 9.45$ and the activation gap is positive ($%
G_{a}\approx 1.53>0$) leading to a stable $\protect\nu =1$ QHE state in the
QW, in agreement the experimental result\protect\cite{wrobel94} in which $%
G_{a}\approx 1.0.$}
\label{Fig2}
\end{figure}

The energy spectra of the spin-split LL subbands calculated within the BS
approach and within the Hartree-Fock approximation, where no correlation
effects are included, are depicted in Fig. 1. Horizontal lines represent the
position of the Fermi level $E_{F}$ obtained from both methods. The
parameters used in calculations are those for sample 1 of Ref.%
\onlinecite{wrobel94}. We see that for these values $r_{0}\approx 0.82$, $%
\omega _{c}/\Omega \approx 26.6$, and $v_{g}/v_{g}^{H}\ \approx 8.5$, the
criterion of validity of the BS approach is fulfilled and, because the
activation gap is negative ($G_{a}\approx -2.9$), any stable $\nu =1$
quantum Hall effect (QHE) state does exist in agreement with experiment.\cite%
{wrobel94} As shown in Fig. 1, in the BS approach, $E_{F}$ is actually the
quasi-Fermi level when $G_{a}<0$.
\begin{figure}[tbp]
\includegraphics*[width=1.0\linewidth]{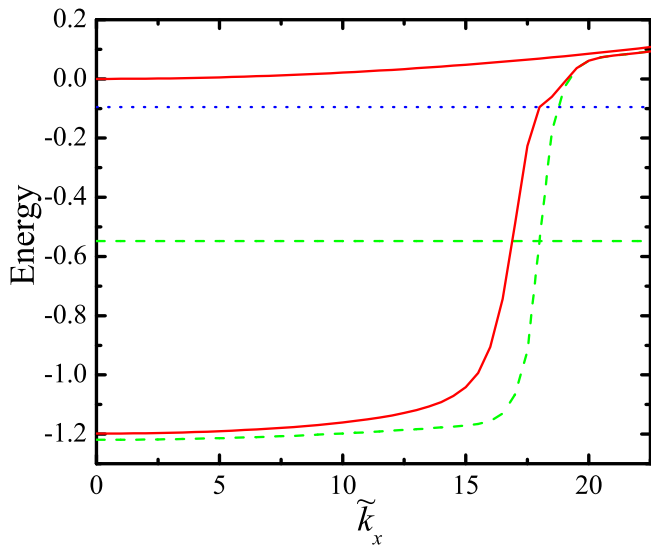}
\caption{(Color on line) Same as in Fig. 2 for the confining frequency $%
\hbar \Omega =0.26$ meV, which corresponds to the estimated threshold for
sample 2 in Ref. \onlinecite{wrobel94}. Here $\protect\omega _{c}/\Omega
\approx 48.5$, where $v_{g}/v_{g}^{H}\approx 16.2,$ and $G_{a}\approx 12.9.$}
\label{Fig3}
\end{figure}

In Figs. 2 and 3 we show the calculation results for the pertinent
parameters of the sample 2 of Ref. \onlinecite{wrobel94}. In Fig. 2 we took
the most probable experimental values of $\hbar \Omega =0.46$ meV and $%
W\approx 0.34$ $\mu $m, and now the magnetic field $B=7.3$ T which leads to $%
r_{0}\approx 0.96$, $\omega _{c}/\Omega \approx 27.4,$ $v_{g}/v_{g}^{H}%
\approx 9.45$. We find that now $G_{a}$ is positive ($G_{a}\approx 1.53$)
and close to the experimental value $G_{a}\approx 1.0$ which corresponds to
the existence of a stable $\nu =1$ QHE state in this sample.\cite{wrobel94}
Figure 3 exhibits the spin-split LL subbands for another confinement
frequency $\hbar \Omega =0.26$ meV (the estimated threshold value in\cite%
{wrobel94} for sample 2). In this case $\omega _{c}/\Omega \approx 48.5$,
where $v_{g}/v_{g}^{H}\approx 16.2,$ and\ $G_{a}\approx 12.9$. Again a
stable the QHE state is predicted to exist.
\begin{figure}[tbp]
\includegraphics*[width=1.0\linewidth]{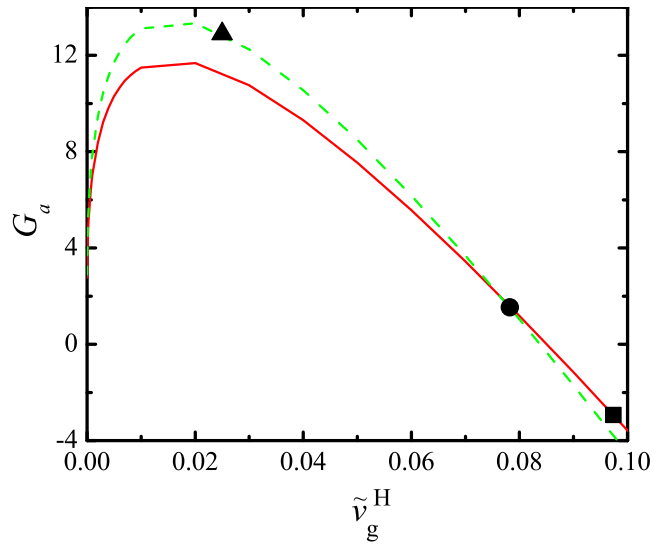}
\caption{(Color on line) Dimensionless activation gap $G_{a}$, or activation
gap $G$ in units of $\left\vert g_{0}\right\vert \protect\mu _{B}B/2$, for
the $\protect\nu =1$ state, as a function of the HA group velocity $\tilde{v}%
_{g}^{H},$ calculated within the BS scheme for $k_{F}\ell _{0}=18.0$ and $%
\protect\delta =10^{-3}.$ The red solid and green dashed curves correspond
to $B=10.0$ and $7.3$ T, respectively. The square mark ($G_{a}\approx -2.94$%
, indicating the collapse of the $\protect\nu =1$ state) corresponds to Fig.
1 parameters. The point marked by the circle (triangle) corresponds to Fig.
2 (3) and it $G_{a}\approx 1.53$ ($12.9$) clearly predicts the existence of
the $\protect\nu =1$ QHE state in this QW sample.}
\label{Fig4}
\end{figure}

The dimensionless activation gap $G_{a}$ is shown in Fig. 4 as a function of
the HA group velocity calculated within the BS approach for $B=10.0$ and $%
7.3 $ T. The points, ($\tilde{v}_{g}^{H},G_{a}$), pertinent to parameters of
Figs. 1-3, are represented by the square, circle and triangle marks. They
indicate the collapse of the $\nu =1$ state (square) or its stability
(circle and triangle) for the QWs in the samples of Ref.\onlinecite{wrobel94}%
.
\begin{figure}[tbp]
\includegraphics*[width=1.0\linewidth]{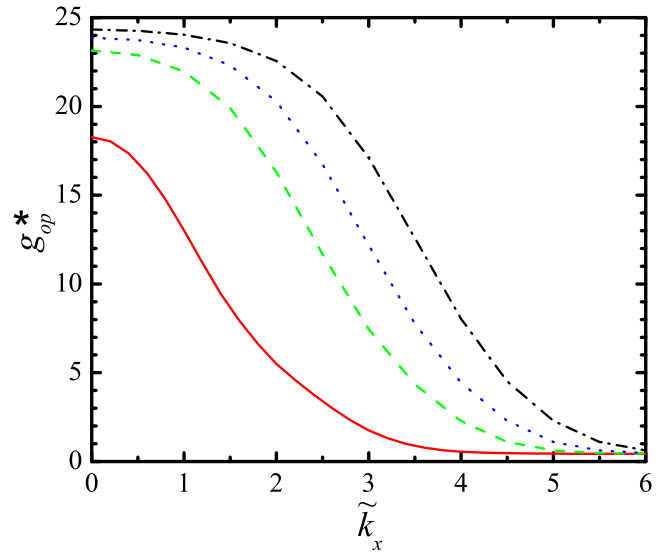}
\caption{(Color on line) Effective \textquotedblleft
optical\textquotedblright $g$-factor $g_{op}^{\ast }$ as a function of $%
\tilde{k}_{x},$ for the experimental conditions of Ref.
\onlinecite{pallecchi02}: $B=14.0$ T, $\hbar \Omega =4.75$ meV. The red
solid, green dashed, blue dotted and black dash-dotted curves are depicted
for $k_{F}\ell _{0}=1.83,$ $3.0$, $3.5$ and $4.0$ corresponding to effective
QW widths of $W\approx 250$, $410$, $480$ and $550$ \AA\ respectively. The
values of the $g_{op}^{\ast }\approx 18.3$, $23.2$ and $23.9$ at $\tilde{k}%
_{x}=0$ for the solid, dashed and dotted curves are very close to the
measured value $g_{op}^{\ast }\approx 21$. The red solid line corresponds to
a QW with linear density $n_{L}=8.5\times 10^{5}$ cm$^{-1}$.\protect\cite%
{pallecchi02}}
\label{Fig5}
\end{figure}

Now we focus on the results, obtained within the BS approach for the QWs,
pertinent to more recent experimental work of Pallechi \textit{et al}.\cite%
{pallecchi02} In Fig. 5 we plot the effective, spatially inhomogeneous,
\textquotedblleft optical\textquotedblright\ $g$-factor $g_{op}^{\ast }$ as
a function of $\tilde{k}_{x}$ for $B=14.0$ T, $\hbar \Omega =4.75$ meV which
corresponds to the weakest lateral confinement in the experiment, at $V_{%
\text{side}}=0$; $\delta =10^{-3}$. The chosen parameters $k_{F}\ell _{0}$
correspond to QW widths smaller than the nominal lithographic width ($\sim
1500$ \AA ), the upper limit for the effective QW width $W$. Typical values
for the $g$-factor at $\tilde{k}_{x}=0$ $g_{op}^{\ast }\approx 18.3$, $23.2$
and $23.9$ from the solid, dashed and dotted curves are rather close to the
measured value ($g_{opt}^{\ast }\approx 21$, see Fig. 5 of\cite{pallecchi02}%
). The solid curve corresponds to the experimental value of the linear
electron density in the QW,\cite{pallecchi02} $n_{L}=8.5\times 10^{5}$ cm$%
^{-1}$. We point out that the existence of the $\nu =1$ state in the QW is
not predicted only for the parameters used for calculation of the
dash-dotted curve. In Fig. 5 the requirements for the applicability of the
BS approach for the QW are well fulfilled; $r_{0}\approx 0.69$, $\omega
_{c}/\Omega \approx 5.1$ and on the average, $v_{g}/v_{g}^{H}\agt1.0$, $%
\tilde{v}_{g}^{H}<1.0.$ In Fig. 6 dimensionless activation gap $G_{a}$ is
plotted as a function of $\tilde{v}_{g}^{H}$ (here it is $\varpropto $ $%
\Omega ^{2}/\omega _{c}^{2}$ ) calculated within the BS approach for $B=14.0$
T. The curves are shown for the same parameters as in Fig. 5. However, in
contrast with Fig. 5, $\Omega $ is now a variable parameter.

\begin{figure}[tbp]
\includegraphics*[width=1.0\linewidth]{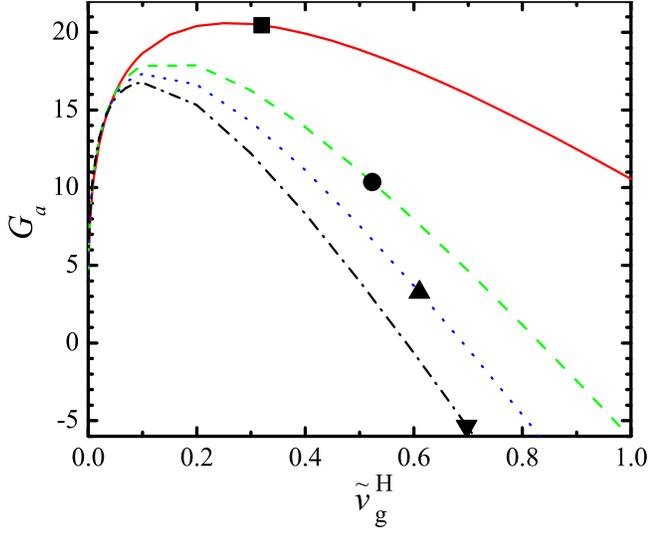}
\caption{(Color on line) Dimensionless activation gap $G_{a}$ as a function
of $\tilde{v}_{g}^{H}$ (or $\Omega ^{2}/\protect\omega _{c}^{2}$ )
calculated within the BS scheme for $B=14.0$ T and $\protect\nu =1$. The red
solid, green dashed, blue dotted and black dash-dotted curves were
calculated for the same values of $k_{F}\ell _{0}$ and $W$ as given in Fig.
5. The square, circle, triangle, inverse-triangle symbols indicated the
values of $G_{a}$ for $\hbar \Omega =4.75$ meV (or $\protect\omega %
_{c}/\Omega \approx 5.1$), used in Fig. 5. It is seen that the
inverse-triangle mark ($G_{a}=-5.35$) implies the collapse of the $\protect%
\nu =1$ state.}
\label{Fig6}
\end{figure}

\section{Collapse of the $\protect\nu =1$ quantum Hall state in the quantum
wire}

We now turn to the important question of collapse of the $\nu =1$ QHE state
in a QW, due to the suppression of exchange-enhanced spin splitting for the
two different theoretical scenarios of the collapse, or the phase
transition, which we propose to understand the experimental findings. In the
first one, there is no change in the QW width at $B_{cr}^{(1)}$, when the
Fermi level, $E_{F}$, reaches the bottom of ($n=0,\sigma =1$) LL. This
scenario is developed here by employing the BS self-consistent approach
(beyond the HFA), as discussed in Sec. II. The second scenario is similar to
one proposed by Kinaret and Lee, where the collapse of the $\nu =1$ state is
caused by the transition to the $\nu =2$ state in the centre of the QW. The
latter state has an effective width two times smaller that the former one if
we ignore the bare $g$-factor; $g_{0}=-0.44$ for GaAs samples. For a more
accurate description of this phase transition some improvements in the study
of Ref. \onlinecite{kinaret90}, discussed partly in Sec. I, are necessary.

\subsection{Collapse of the state within the second scenario}

We will not restrict ourselves, in this subsection, to the limit $\Omega
^{2}/\omega _{c}^{2}\ll 1$, but consider also the important case $\Omega
/\omega _{c}\agt1$, by using Eqs. (\ref{1}), (\ref{2}) and other general
formulas of Sec. II valid for arbitrary $\Omega /\omega _{c}$. Within the
restricted HFA model the single-particle energies of the two lowest
spin-split LLs ($n=0$, $\sigma =\pm 1$) are written as%
\begin{align}
E_{0,k_{x},\pm 1}^{F}& =\frac{\hbar \tilde{\omega}}{2}\mp \frac{|g_{0}|\mu
_{B}B}{2}+\frac{\hbar ^{2}}{2m^{\ast }}(\frac{\Omega }{\tilde{\omega}}%
)^{2}k_{x}^{2}  \notag \\
& -\frac{e^{2}}{\pi \varepsilon }\int_{k_{x}-k_{F0}^{(\pm
1)}}^{k_{x}+k_{F0}^{(\pm 1)}}dxM_{a}(0,x),  \label{14}
\end{align}%
where the last term corresponds to the exchange interaction. The function $%
M_{a}(0,q_{x})=2^{-1}\exp [-(2a^{2}-1)q_{x}^{2}\tilde{\ell}%
^{2}/4]K_{0}(q_{x}^{2}\tilde{\ell}^{2}/4)$ for $\Omega ^{2}/\omega
_{c}^{2}\rightarrow 0$ (correspondingly, $a\rightarrow 1$) coincides with $%
M(0,q_{x})$, and $k_{F0}^{(\pm 1)}$ is the Fermi wave vector of the
spin-split level. Note that for the $\nu =1$ state of the QW, $%
k_{F0}^{(-1)}=0$ as only the ($n=0$,$\sigma =+1$) LL is occupied, the
exchange contribution in $E_{0,k_{x},-1}^{F}$ vanishes. The exchange
interaction, given in Eq. (\ref{14}), coincides with Eq. (10) of Ref.%
\onlinecite{kinaret90} only in the limit $\Omega ^{2}/\omega
_{c}^{2}\rightarrow 0$ and it is essentially different for $\Omega /\omega
_{c}\agt1$, in particular, due to the fact that $a\neq 1$.

Integrating Eq. (\ref{14}) over $k_{x}$ from $-k_{F0}^{(\pm 1)}$ to $%
k_{F0}^{(\pm 1)}$, after taking half of the exchange term to avoid double
counting, and then summing the result for these two levels we arrive to the
expression for the total energy of the Q1DES in QW, per unit of length as%
\begin{align}
E^{F,tot}(\lambda )& =\frac{\hbar \tilde{\omega}}{2}n_{L}-|g_{0}|\mu
_{B}B\lambda +\frac{\pi ^{2}\hbar ^{2}}{6m^{\ast }}(\frac{\Omega }{\tilde{%
\omega}})^{2}\sum_{p=0}^{1}[\frac{n_{L}}{2}  \notag \\
& +(-1)^{p}\lambda ]^{3}-\frac{e^{2}}{2\pi ^{2}\varepsilon }%
\sum_{p=0}^{1}\int_{0}^{\pi \lbrack n_{L}/2+(-1)^{p}\lambda ]}dk_{x}  \notag
\\
& \times \int_{k_{x}-\pi \lbrack n_{L}/2+(-1)^{p}\lambda ]}^{k_{x}+\pi
\lbrack n_{L}/2+(-1)^{p}\lambda ]}dxM_{a}(0,x),  \label{15}
\end{align}%
where $\lambda $ is the linear density asymmetry between the spin-split
levels and $k_{F0}^{(\pm 1)}=\pi \lbrack n_{L}/2\pm \lambda ]$. Notice, the $%
p$-term of last sum of Eq. (\ref{15}) cannot be reduced to Eqs. (13) and
(14) of Ref. \onlinecite{kinaret90} for any finite $\Omega /\omega _{c}$.
However, for $\Omega ^{2}/\omega _{c}^{2}\ll 1$ the relative difference
between them becomes negligible. It is implicit in Eq. (\ref{15}) that if
one spin-split LL is occupied then the condition (i) $%
E_{0,k_{F0}^{(+1)},+1}^{F}<\epsilon _{0,0,-1}$ should be satisfied. On the
other hand, if both spin-split LLs are occupied then the following condition
of thermodynamical stability (ii) $%
E_{0,k_{F0}^{(+1)},+1}^{F}=E_{0,k_{F0}^{(-1)},-1}^{F}$ should be satisfied.
We note that the conditions (i) and (ii) are actually fulfilled only for
one, two or three\ values of $\lambda $ within the range $n_{L}/2\geq
\lambda \geq 0$ depending on $\omega _{c}$, $\Omega $ and other parameters.
It does mean that each one of the three curves of Fig. 2 in Ref. \onlinecite{%
kinaret90} (plotted for three values of $n_{L}$) are really reduced, due to
necessary conditions (i) and (ii), to three points corresponding to $\lambda
=0,$ $n_{L}/2$, and a third $\lambda $ which has a more specific value for
each curve.

By applying Eqs. (\ref{14}), (\ref{15}) to GaAs-based QWs of Refs. %
\onlinecite{wrobel94} and \onlinecite{pallecchi02}, with fixed $n_{L}$, we
obtain that for any $\lambda \in \lbrack 0,n_{L}/2]$ the results for the
critical magnetic field $B_{cr}^{(2),F}$ for the actual value of $%
|g_{0}|=0.44$ are very close to results for $g_{0}=0$. Indeed, in Eq. (\ref%
{15}), $|g_{0}|\mu _{B}B\ll e^{2}/\varepsilon \tilde{\ell}$. So in very good
approximation we further assume that $g_{0}=0$.

Our analysis shows that the state with the lowest total energy corresponds
to $\lambda =0$ (equally occupied spin-split LLs: $\nu =2$ state) or to $%
\lambda =n_{L}/2$ (one occupied spin-split LL: $\nu =1$ state). Then to
calculate $B_{cr}^{(2),F}$, within the second scenario where the $\nu =1$
state with width $W$ ($k_{F0}^{(1)}=k_{F}$, $k_{F0}^{(-1)}=0$) collapses to
the $\nu =2$ state with width $W/2$ ($k_{F0}^{(\pm 1)}=k_{F}/2$), we need to
solve the equation%
\begin{equation}
\Delta E^{F,tot}(B,\Omega ;n_{L})\equiv E^{F,tot}(n_{L}/2)-E^{F,tot}(0)=0.
\label{16}
\end{equation}%
For fixed $n_{L}$, from Eq. (\ref{16}), we obtain the critical curve $\Omega
(B_{cr}^{(2),F})$. We point out that the latter curve is different from the
equivalent critical curve $\Omega (B_{cr}^{(2),KL})$, calculated within the
Kinaret-Lee model due to the inappropriate calculation of the exchange
interaction term, as outlined above. In contrast with the first scenario
discussed in the previous section where, at $B_{cr}^{(1)}$, there is no
change of the effective QW width, now the QW width drops sharply by a factor
$2$ at a certain $B_{cr}^{(2),F}$. As a consequence, this strong
redistribution of the charge density in the QW compels us to add the Hartree
interaction that results in the direct interaction term,

\begin{equation}
\Delta E^{H}=\frac{4e^{2}}{\pi ^{2}\varepsilon a^{2}\tilde{\ell}^{2}}%
\int_{0}^{\infty }\frac{du}{u^{3}}e^{-u^{2}/2}[1-\cos ^{2}(\frac{ak_{F}%
\tilde{\ell}u}{2})]^{2},  \label{17}
\end{equation}%
modifying the Eq. (\ref{16}) and leading to the correct expression for the
critical point in the HFA, as%
\begin{equation}
\Delta E^{HF,tot}(B,\Omega ;n_{L})\equiv \Delta E^{F,tot}(B,\Omega
;n_{L})-\Delta E^{H}=0.  \label{18}
\end{equation}%
We observe that because $\Delta E^{H}>0$, for any finite $B$, it follows
from Eq. (\ref{18}) that, in the HFA, the Hartree interaction contributes to
make the total energy of the $\nu =1$ state lower than that of the $\nu =2$
QW state. We will see that for $\Omega /\omega _{c}\alt1$ and $n_{L},\Omega $
fixed, the critical magnetic field $B_{cr}^{(2)}$, calculated from Eq. (\ref%
{18}), is very different from $B_{cr}^{(2),F}$, calculated by neglecting the
Hartree interaction. It is easy to see that for $\Omega /\omega
_{c}\rightarrow \infty $ the difference between Eqs. (\ref{16}) and (\ref{18}%
) becomes negligible, and $B_{cr}^{(2),F}\simeq B_{cr}^{(2)}$.

\begin{figure}[tbp]
\includegraphics*[width=1.0\linewidth]{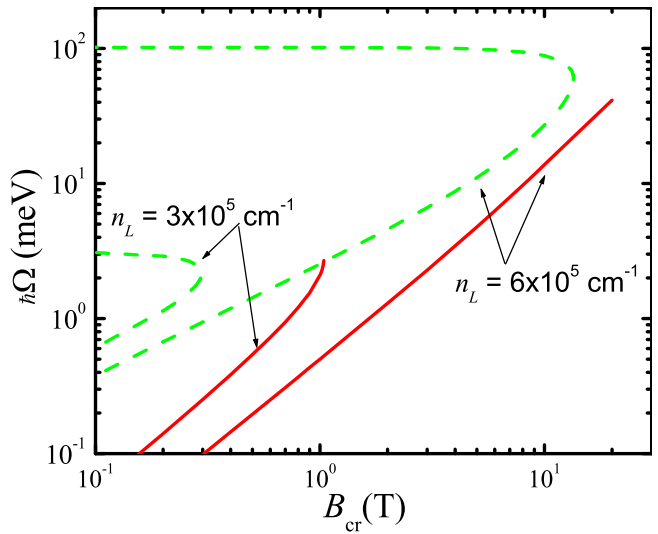}
\caption{(Color on line) Critical curves $\Omega $ vs $B_{cr}$ for the
collapse ($G_{a}=0$) of the $\protect\nu =1$ state in the GaAs-based QWs,
for two linear densities. The first scenario $\Omega =\Omega (B_{cr}^{(1)})$%
, shown by the red solid curves, is calculated within the BS approach. The
second scenario $\Omega =\Omega (B_{cr}^{(2)})$, indicated by the green
dashed curves, is evaluated in the HFA from Eq. (\protect\ref{18}).}
\label{Fig7}
\end{figure}

\subsection{Phase diagram for the $\protect\nu =1$ state collapse}

The phase diagrams for the collapse of the $\nu =1$ state of the interacting
Q1DES, laterally confined by a parabolic potential with characteristic
frequency $\Omega $, that follow from the first and the second scenarios,
are plotted on a logarithmic scale in Fig. 7. The horizontal axis represents
the critical magnetic field $B_{cr}$ at which the activation gap is
suppressed, while the vertical axis is the confinement frequency on a scale
compatible with parameters for GaAs-based samples. The red solid curves
represent the suppression of the activation gap, driven by
exchange-correlation effects in the QW that are calculated for the proposed
first scenario within the BS approach. The green dashed lines are obtained
from the solution of Eq. (\ref{18}) and corresponds to the second scenario
where the phase transition due to equal population of both spin-split LLs
occurs at $B_{cr}^{(2)}$.

\begin{figure}[tbp]
\includegraphics*[width=1.0\linewidth]{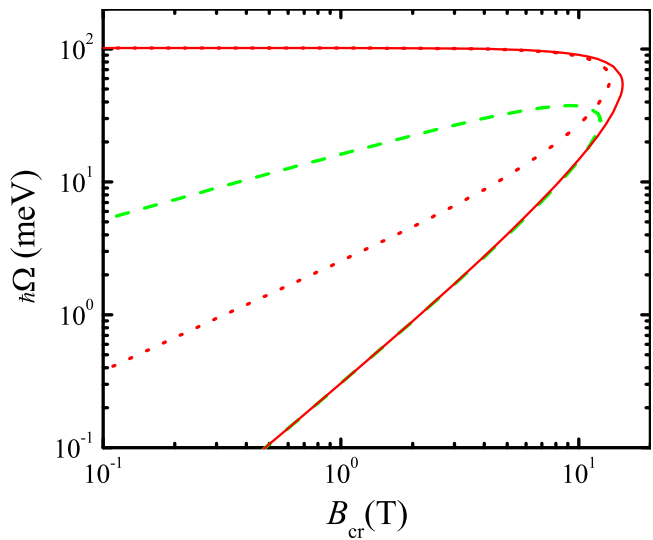}
\caption{(Color on line) Comparison of the second scenario result for the
critical curve $\Omega =\Omega (B_{cr}^{(2)})$ plotted by the red solid
curve (from Eq. (\protect\ref{18})) with the result from the Kinaret and Lee
scenario $\Omega =\Omega (B_{cr}^{(2),KL})$ plotted by the green dashed
curve (from their Eqs. (12)-(14) in which the Hartree interaction is not
taken into account\protect\cite{kinaret90}). The result of the second
scenario with the Hartree interaction discarded, $\Omega =\Omega
(B_{cr}^{(2),F})$, is plotted by the red dotted curve. The linear density
for a QW is $n_{L}=6\times 10^{5}$ cm$^{-1}$.}
\label{Fig8}
\end{figure}

In Fig. 8 we compare the critical curve $\Omega $ vs $B_{cr}$ (red solid
line), calculated within the HFA [Eq. (\ref{18}), for the second scenario
with the result obtained from the solution of Eq. (\ref{16}), or Eq. (\ref%
{18}) where the Hartree interaction contribution $\Delta E^{H}$ is excluded
(red dotted line). We also show the critical curve $\Omega (B_{cr}^{(2),KL})$
calculated for the Kinaret-Lee model, according to the Eqs. (12)-(14) of Ref.%
\onlinecite{kinaret90} (green dashed line).

Critical curves $\Omega (B_{cr})$ are depicted in Fig. 9 in a linear-scale
plot for the experimental conditions of Ref. \onlinecite{pallecchi02}. The
curves split the region $\nu =2$ on the left side from the region $\nu =1$
on the right side of the phase diagram. The red solid curve represents $%
\Omega (B_{cr}^{(1)})$\ for the first scenario calculated in the BS approach
and the red dot-dashed line shows $\Omega (B_{cr}^{(1),HF})$ in the HFA. The
comparison between these curves indicates the role of electron correlations
in the QW system for the first scenario. The green dashed curve indicates $%
\Omega (B_{cr}^{(2)})$ for the second scenario in the HFA (from Eq. (\ref{18}%
)) and $\Omega (B_{cr}^{(2),KL})$ is depicted by the blue dotted curve. Our
results for $\Omega (B_{cr}^{(2),F})$ (calculated from Eq. (\ref{16}), but
not shown in Fig. 9) coincide with those from Ref. \onlinecite{kinaret90},
because $\omega _{c}^{2}/\Omega ^{2}\gg 1$ for the curves depicted in Fig. 9.

\begin{figure}[tbp]
\includegraphics*[width=1.0\linewidth]{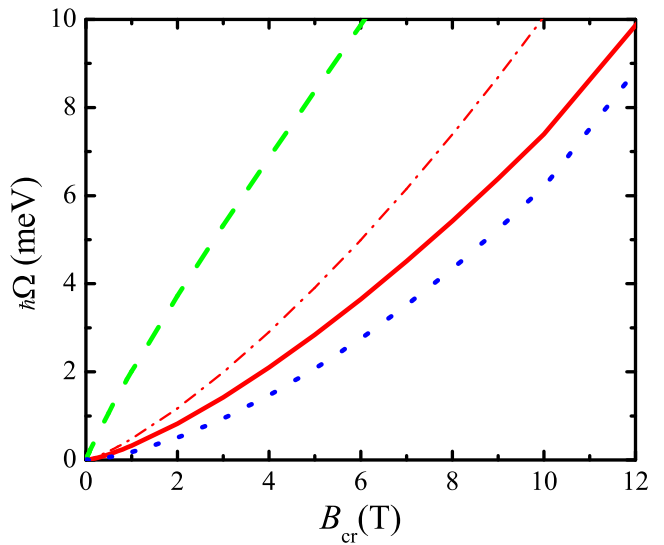}
\caption{(Color on line) Linear scale plot of the critical curves for the
experimental conditions of Ref. \onlinecite{pallecchi02}, with $%
n_{L}=8.5\times 10^{5}\;$cm$^{-1}$. The first scenario result, $\Omega
=\Omega (B_{cr}^{(1)})$, calculated in the BS scheme, is given by the red
solid curve and the red dot-dashed curve plots, $\Omega =\Omega
(B_{cr}^{(1),HF})$, the same curve, however, with neglected correlations,
i.e., in the HFA. One can see clearly, by comparing these curves, the role
of electron correlations. The green dashed line represents the second
scenario result, $\Omega =\Omega (B_{cr}^{(2)})$, in the HFA. The result of
the Ref. \onlinecite{kinaret90} model, $\Omega =\Omega (B_{cr}^{(2),KL})$,
is denoted by the blue dotted curve; in Fig. 9 it will practically coincide
with our result, $\Omega =\Omega (B_{cr}^{(2),F})$, for omitted direct
interaction in the second scenario. It can be seen that only the red solid
curve, obtained within the first scenario, can explain the observed critical
magnetic fields in which the collapse of the $\protect\nu =1$ state occurs
for different values of\protect\cite{pallecchi02} $\Omega $ as the blue
dotted curve should be discarded.}
\label{Fig9}
\end{figure}

Various experimental consequences of our theoretical analysis are now
discussed. For the lowest confinement frequency $\hbar \Omega =4.75$ meV,
taken from the experiment of Ref. \onlinecite{pallecchi02} for the side gate
voltage, which controls the lateral confinement, $V_{\text{side}}=0$, the
predicted critical magnetic fields from curves of Fig. 9 are $%
B_{cr}^{(1)}=7.26$ T, and $B_{cr}^{(2)}=2.65$ T, whereas the experimental
value is $B_{cr}=7$ T. The close agreement is a clear evidence that the
first scenario is realized in this case. Note that here $B_{cr}^{(2),F}%
\approx B_{cr}^{(2),KL}=8.47$ T and $B_{cr}^{(1),HF}=5.73$ T for $\hbar
\Omega =4.75$ meV. For the largest confinement frequency, $\hbar \Omega
\approx 7.0$ meV, obtained for the largest negative $V_{\text{side}}$ in
Ref. \onlinecite{pallecchi02}, we have $B_{cr}^{(1)}=9.6$ T and $%
B_{cr}^{(2)}=4.1$ T that can be compared with the experimental value $%
B_{cr}=10$ T. This allows us to conclude that the first scenario is again
realized. Note that for the same confining frequency it follows that $%
B_{cr}^{(1),HF}=7.7$ T and $B_{cr}^{(2),F}\approx B_{cr}^{(2),KL}=10.8$ T.
That is surprising that for some specific values of the confinement
frequency, the results taking into account only the exchange interaction can
lead to $B_{cr}$ close to the experimental ones, even though, from the
theoretical point of view, we have shown that the Hartree direct term is
quite essential because the strong redistribution of the electron density in
this scenario.

Now we continue our discussion by analyzing the realization of two scenarios
for the sample parameters of the older experiment.\cite{wrobel94} In Fig.
10, we depict the critical curves $\Omega (B_{cr}^{(1)})$ and $\Omega
(B_{cr}^{(2)})$ by red solid lines and green dashed lines respectively. We
plot also for comparison $\Omega (B_{cr}^{(2),KL})$, obtained within the
Kinaret-Lee model by the blue dotted lines; there is no noticeable
difference between $\Omega (B_{cr}^{(2),F})$ and $\Omega (B_{cr}^{(2),KL})$
in this case. For parameters of sample 1, used in Fig. 1 ($\hbar \Omega =0.65
$ meV), we observe that $B_{cr}^{(1)}=10.8$ T (for $g_{0}=-0.44$ it should
be $B_{cr}^{(1)}=10.4$ T) and $B_{cr}^{(2)}=1.27$ T. The other critical
fields $B_{cr}^{(2),KL}\approx B_{cr}^{(2),F}=20.0$ T. These results support
the occurrence of the first scenario, because the collapse of the $\nu =1$
state was not observed even though it was expected to occur at $B=10$ T.\cite%
{wrobel94} We see again how important is the Hartree contribution for the
stability of the $\nu =1$ state in the second scenario. Furthermore, for the
parameters of sample 2 of Ref. \onlinecite{wrobel94}, used in Fig. 2 ($\hbar
\Omega =0.46$ meV), we find $B_{cr}^{(1)}=7.3$ T (for finite $g_{0}=-0.44$,
it should be $B_{cr}^{(1)}=7.0$ T), $B_{cr}^{(2)}=0.82$ T, and $%
B_{cr}^{(2),KL}\approx B_{cr}^{(2),F}=13.1$ T. In addition, for the
parameters used in Fig. 3 ($\hbar \Omega =0.26$ meV) of the same sample, we
obtain $B_{cr}^{(1)}=4.8$ T (notice, for finite $g_{0}=-0.44$, it should be $%
B_{cr}^{(1)}=4.6$ T), $B_{cr}^{(2)}=0.46$ T, and $B_{cr}^{(2),KL}\approx
B_{cr}^{(2),F}=9.2$ T. As for the fixed confinement frequencies in Fig. 10,
we concluded that the theoretical results indicate again the occurrence of
the first scenario, because the observed $\nu =1$ QHE state persists with
the centre of the plateau at $B=7.3$ T.\cite{wrobel94}

\begin{figure}[tbp]
\includegraphics*[width=1.0\linewidth]{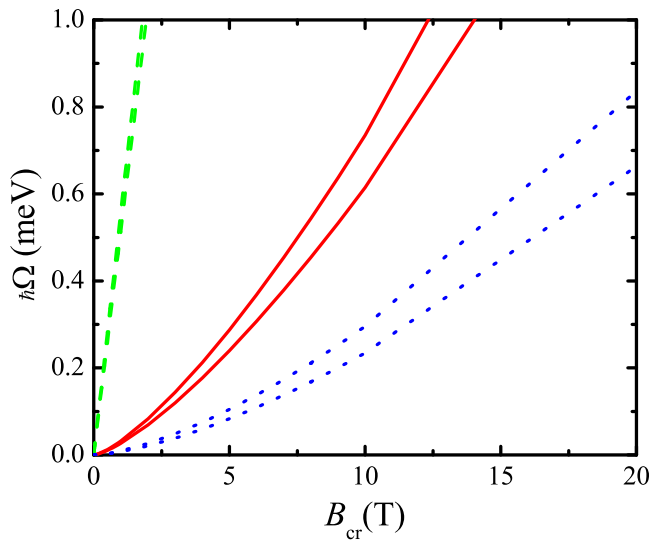}
\caption{(Color on line) Same critical curves, as in Fig. 9, for the sample
1 (sample 2) quantum wire of Ref. \onlinecite{wrobel94} with $n_{L}=7.0$ $%
(6.0)\times 10^{6}\;$cm$^{-1}$ plotted by the lower (upper) the red solid
line, $\Omega =\Omega (B_{cr}^{(1)})$, the green dashed curve, $\Omega
=\Omega (B_{cr}^{(2)})$, and the blue dotted line, $\Omega =\Omega
(B_{cr}^{(2),KL})$. Only the first scenario results explain well the
experimental observations\protect\cite{wrobel94} (see the text).}
\label{Fig10}
\end{figure}

\section{Conclusions}

We have studied in this work two scenarios for phase transitions leading to
the collapse of the $\nu =1$ quantum Hall state in a QW due to the
suppression of exchange-enhanced activation $g$-factor. In the first
scenario the collapse of the activation gap $G_{a}$ (as well as the $g$%
-factor) occurs at $B_{cr}^{(1)}$ without any finite redistribution of the
charge density in the QW; we have obtained it for a still strong
exchange-enhanced optical $g$-factor $g_{op}^{\ast }(0)$ at the centre of
the QW (see Fig. 5),\ which is in reasonable agreement with the experimental
results.\cite{pallecchi02} Within the second scenario of the collapse of the
quantum Hall state, there is a strong decrease of the electron width $W$ of
the QW at $B_{cr}^{(2)}$. In this case $g_{op}^{\ast }(k_{x})$ drops to zero
for any $k_{x}$. Because the electron density is strongly redistributed in a
narrow region, the Hartree term of the total energy plays essential role and
must be included in the calculations.

We call the attention for an important point coming from our theoretical
investigations. In the second scenario, it follows (see Fig. 8, for
instance) that, for a given $n_{L}$, there is $\Omega _{0}$ such that for $%
\Omega >\Omega _{0}$ and for any $B$, in particular, for $B\rightarrow 0$,
the $\nu =1$ state should be stable. Furthermore, $\Omega $ can be chosen
sufficiently large so that the parameter $e^{2}/(\varepsilon \tilde{\ell}%
\hbar \tilde{\omega})\ $would be extremely small. However, this contradicts
the Lieb-Mattis theorem \cite{liebmattis62} that assures that the ground
state of 1D many-body system is demagnetized. \cite{starykh99,sushkov03}
This result reinforces the role of correlations for weak magnetic fields.

Our study, using the extended BS approach, demonstrated the importance to
take into account correlation effects, due to edge states screening, for the
dependence of the LLs on the position nearby the edges.

We have compared the theoretical results in both scenarios with
experiments performed by two different
groups.\cite{wrobel94,pallecchi02} Even though a direct comparison
with experiments should be difficult, due to different samples and
accuracy of essential parameters, our overall conclusion is that
the first scenario is most favorable to be realized in QWs.

\begin{acknowledgments}
This work was supported in part by Funda\c{c}\~{a}o de Amparo \`{a} Pesquisa
do Estado de S\~{a}o Paulo (FAPESP). One of the authors (S.S.) is grateful
to Coordena\c{c}\~{a}o de Aperfei\c{c}oamento de Pessoal de N\'{\i}vel
Superior (CAPES) for a doctoral fellowship and to Universidade do Amazonas
for a leave of absence. N.S acknowledges Conselho Nacional de
Desenvolvimento Cient\'{\i}fico e Tecnol\'{o}gico (CNPq) for a research
fellowship.
\end{acknowledgments}

\end{document}